\documentclass[pop,aps,floats,preprint,graphics,amsmath,amssymb,superscriptaddress,showpacs,pre]{revtex4-1}
\usepackage{graphics,graphicx,color}
\usepackage[utf8]{inputenc}

\begin{document}

\title{On the relation between non-exponential Scrape Off Layer profiles and the dynamics of filaments}

\author{F. Militello} 
\affiliation{CCFE, Culham Science Centre, Abingdon, Oxon, OX14 3DB, UK}
\author{J.T. Omotani}
\affiliation{Department of Physics, Chalmers University of Technology, SE-412 96 Göteborg, Sweden}

\begin{abstract}
A theoretical framework is developed to clarify the relation between the profiles of density and temperature in the Scrape Off Layer (SOL) with the fluctuations (filaments) that generate them. The framework is based on the dynamics of independent filaments and on their statistical behaviour and can be used to rigorously understand the mechanisms that lead to the non-exponential nature of the radial SOL profiles as well as the increase of the relative fluctuation amplitude in the far SOL. Several models for the dynamics of the filaments, which can be applied to the framework, are derived and discussed for the purpose of identifying how different assumptions lead to the emergence of features in the profiles. It is found that multiple alternative models can explain the observations, thus motivating more stringent and focused experimental analysis. In particular, radially accelerating filaments, less efficient parallel exhaust and also a statistical distribution of the velocity of the filaments can all contribute to induce flatter profiles in the far SOL. A quite general result is the resiliency of the non-exponential nature of the profiles. At the same time, several of the models discussed can also capture the increase of the relative fluctuation amplitude observed in the far SOL. It is also shown that several scenarios are compatible with the broadening of the SOL, which could be caused by charge exchange interactions with neutral particles or by a significant radial acceleration of the filaments.   
\end{abstract}

\maketitle

\section{Introduction} \label{sec0}

The exhaust of power and particles in experimental magnetic fusion devices determines the level of interaction between the plasma and the material surfaces \cite{Loarte2007}. Next generation reactor relevant machines are expected to operate in conditions where such an interaction, unless properly controlled, might become extremely problematic for the lifetime of the plasma facing components. The exhaust occurs through a narrow region of plasma surrounding the magnetic separatrix called the Scrape Off Layer (SOL), where the field lines are open and connected to solid surfaces. 

Experiments in the last 20 years showed that, rather universally across machines, the midplane density and electron temperature profiles tend to flatten at a certain distance from the separatrix, in the so called far SOL \cite{Asakura1997,LaBombard1997,LaBombard2001,Lipschultz2002,Whyte2005,Lipschultz2005,Garcia2007,Garcia2007b,Carralero2014,Militello2016}. This has practical consequences since broad profiles redirect the plasma towards the first wall rather than towards the divertor components, which are specifically designed to sustain the large fluxes associated with the exhaust \cite{Umansky1998,LaBombard2000}. The non-exponential nature of the profiles, which we call \textit{flattening} in the rest of the paper, led to the distinction between a near SOL, close to the separatrix, where the gradients are steep, and a far SOL, with slowly varying profiles and further out towards the wall \cite{LaBombard2001}. While this terminology originated in particle transport studies and density profiles, it extended also to the behaviour of the temperature. Another feature that appears to be universal to all the measurements is the response of the density profiles to increasing fuelling levels. Both in the near and far SOL, the decay length becomes longer at a higher fuelling level, but in the latter the change is much stronger, so that the two regions respond in a different way to the main plasma conditions. 

\begin{figure}
\includegraphics[height=8cm,width=8cm, angle=0]{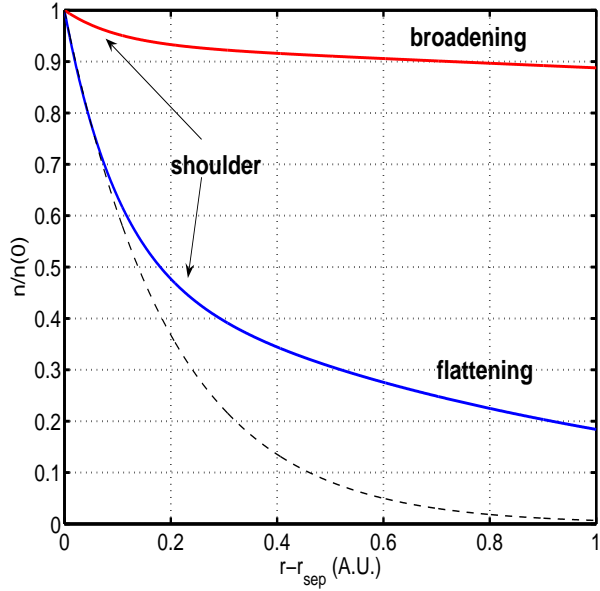}
\caption{Schematic representation of typical SOL profiles at low (blue curve) and high (red curve) fuelling. The dashed line represents a pure exponential decay and its used to show the effect of the flattening. The far SOL extends outwards from the shoulder position.}
\label{fig1}
\end{figure}

At the same time, as the fuelling level increases, the boundary between near and far SOL, called the \textit{shoulder}, moves closer to the separatrix \cite{LaBombard2001,Militello2016} and the near SOL shrinks accordingly. To avoid ambiguity, we introduce a practical definition of the shoulder as the position of maximum curvature of the logarithm of the profile (for an exponential profile this quantity would be zero everywhere and hence the shoulder would be undefined). At high fuelling levels, the far SOL is almost flat and pervades most of the open field line region (if not all). We call this regime density \textit{broadening} in order to avoid confusion with the density \textit{flattening} of the far SOL, see Fig.\ref{fig1}. It is unclear whether the broadening occurs as a  transition or is simply a gradual increase of the flattening. For the purpose of this paper, we will treat the two phenomena as independent and possibly triggered by different physics.    

Several experiments reported that the density profiles are also affected by the plasma current \cite{McCormick1992,Rudakov2005,Garcia2007c,Militello2016}. In particular, higher currents correspond to steeper gradients globally and can even prevent the broadening at comparable line averaged density \cite{Militello2016}. It is possible that this effect is related to a reduction of the connection length, as the profiles significantly steepen in the wall shadow \cite{Lipschultz2002,Rudakov2005,Garcia2007}, where the field lines impinge on limiters rather than the divertor, with a step change reduction in the connection length. These results suggest that parallel physics plays a crucial role in determining the shape of the profiles. 

Interestingly, the electron temperature profiles show a certain degree of flattening but they typically do not respond to fuelling scans \cite{LaBombard1997,Lipschultz2002,Whyte2005}. This suggests that the electron temperature is not a crucial element for the density broadening and also that the temperature (heat) and density exhaust might be regulated by different mechanisms \cite{Militello2016}. A further consequence of this observation is that increased ionisation is generally unlikely to generate and sustain the broadened profiles as, if this was the case, the electron temperature should decrease. Indeed, the additional ionisation would remove energy from the colliding electrons, thus cooling them down for identical power crossing the separatrix. It is worthwhile noticing that the majority, but not all, literature reports temperature profiles insensitive to the line averaged density or the amount of fuelling. An interesting exception is represented by the results in \cite{LaBombard2001} where the temperature profiles also display a clear broadening (and an absolute decrease). Investigations of the ion temperature profile are few \cite{Elmore2012,Allan2016} and, to our knowledge, do not allow a proper characterisation of the profiles in the far SOL. This is unfortunate, since the energy of the ions impinging on the material surfaces determines their erosion.

The flattening of the target heat flux in the far SOL was recently reported in limiter discharges on JET \cite{Arnoux2013} and COMPASS \cite{Horacek2014}. Also in this case, the profiles are steeper close to the separatrix but less so in the far SOL, although in this case they are measured at the limiter surface and not at the midplane. Unlike the density, but like the temperature, the broadening of the profile does not seem to be present in any of the results reported. This might be expected, since the heat flux is largely determined by the electron temperature (with a conductive approximation). 

During L-mode operations, both density flattening and broadening occur routinely in several machines. The first work recognising the operational importance and the implications of flat profiles was Ref.\cite{Umansky1998}. Even earlier, many authors reported similar observations in ASDEX \cite{McCormick1992}, JT-60U \cite{Asakura1997,Asakura1999} ALCATOR C-MOD \cite{LaBombard1997,LaBombard2000,LaBombard2001,Lipschultz2002,Lipschultz2005}, DIII-D \cite{Lipschultz2005,Whyte2005,Rudakov2005}, TCV \cite{Garcia2007,Garcia2007b}, JET and ASDEX-U \cite{,Carralero2014} and finally MAST \cite{Militello2016}. This suggests that this phenomenon is rather universal, robust and not significantly affected by divertor design or geometry. The mechanism behind the flattening and broadening remains elusive, although it was proposed to be related to MARFEs \cite{Asakura1997}, SOL ionisation sources \cite{LaBombard2000}, detachment \cite{Carralero2014}, changes in the perpendicular transport due to divertor collisionality \cite{Carralero2014} or modifications of the dynamics of the filaments \cite{Militello2012}. The phenomenology of the L-mode profiles seems to be reproduced also in the inter-ELM phase of H-mode, although only few results are available in literature \cite{LaBombard1997,Boedo2001,Lipschultz2002,Muller2015}.  

It is important to remark that overwhelming evidence shows that transport in the far SOL is dominated by filamentary structures erupting from the main plasma. Visual cameras \cite{Kirk2006,Dudson2008,BenAyed2009}, gas puff imaging \cite{Zweben2002,Terry2003,Myra2006,Garcia2013} and upstream Langmuir probes \cite{LaBombard2001,Boedo2001,Rudakov2002,Boedo2003,Garcia2007b,Garcia2007c,Militello2013} show the presence of large fluctuations of the thermodynamic quantities in the SOL, especially far from the separatrix. It is therefore the time average over these structures that generate the SOL profiles, which are not equilibria in the proper sense. 

A lot of work has been carried out to characterise the SOL fluctuations, both from an experimental (see \cite{DIppolito2011} for a review) and a theoretical point of view \cite{Krasheninnikov2001,Garcia2004,Myra2006b,Militello2012,Militello2013b,Easy2014,Omotani2015,Easy2016,Militello2013,Militello2016b,Ricci2014,Tamain2014}. Of particular interest is the observation that the turbulence seems to have a universal behaviour in the SOL, which can be captured by describing the statistics of the fluctuations through a gamma distribution \cite{Graves2005}, a result explained also by a recent theoretical model \cite{Garcia2012}. Also, results in literature suggest that the non-Gaussian behaviour of the fluctuations is more evident in the far SOL than in the near SOL and everywhere in profiles after the broadening \cite{Graves2005,Militello2013}. In addition, the relative amplitude of the fluctuations (measured as the standard deviation divided by the mean) tends to increase with the radial position \cite{Terry2003,Zweben2004,Zweben2015}, although this increase can be barely visible in certain instances \cite{Garcia2007}. Using gas puff imaging, both the waiting times of the filaments (i.e. the inverse of the average generation rate) and their amplitude were shown to be exponentially distributed \cite{Garcia2013,Garcia2013b}. Recently, this was also confirmed using wall mounted Langmuir probes at JET \cite{Walkden2016}. Finally, a single filament generates a midplane Langmuir probe signal that has a characteristic double exponential shape \cite{Boedo2001,Rudakov2002,Antar2003,Garcia2006a,Militello2013}, which can be traced to its radial profile \cite{Garcia2006}. 

In this paper, the emergence of the features of the SOL profiles is interpreted with the underlying changes occurring in the filamentary dynamics. In particular, we discuss in detail and extend a recently developed theoretical framework \cite{Militello2016c} which relates the statistics of the filaments to the shape of the SOL profiles. The framework provides a rigorous and general basis for testing theoretical and experimental interpretations of the mechanisms determining the profile response to changes in the filaments' behaviour. A large part of the paper is devoted to the description of a number of first principle models addressing different aspects of the filament dynamics. These models are then implemented in the framework to determine how the profiles respond to different effects (e.g. accelerating filaments, reduced parallel transport). 

\section{Theoretical Framework} \label{sec1}

In this Section we present a summary of the formulation of the theoretical framework derived in \cite{Militello2016c}. While the general framework is relatively simple and relies on few assumptions, analytic solutions associated with filamentary models can be obtained only in a limited number of simplified cases. However, these solutions give insight into the profile generation mechanisms, and are discussed in the following subsections. A complete picture requires a numerical solution.  

\subsection{Definitions} \label{sec1.0}

At this point, it is useful to introduce a few important definitions (in italics). We define a \textit{profile} the function representing the ensemble of the time averages of a thermodynamic field taken at different radial positions (for a given a toroidal and poloidal angle). A profile in the SOL (and in the core) does not give information on the fluctuations due to turbulence, but it captures their collective effect. As such, the profile is a purely statistical quantity. The \textit{background} is the environment in which the fluctuations propagate. It is generated by the steady plasma leakage out of separatrix, which could be caused by drifts or collisional diffusion. Also residual plasma left by travelling fluctuations can contribute to it, thus making the background time dependant. Importantly, the background could be quite different from the profile if the fluctuations are comparable in amplitude with the latter. Neither the background nor the profile is a proper plasma \textit{equilibrium}, defined as a time independent state that, if unstable, is the source of \textit{local fluctuations}. On the other hand, the formation of the profile is to a large extent contributed to by \textit{non local fluctuations} travelling through the background but generated somewhere else, so that the existence (or the lack) of an equilibrium is not crucial. In the core, profile, background and equilibrium coincide and most of the fluctuations are local. In the SOL, the situation is more complicated, as the background is different from the profile (especially in the far SOL), most of the fluctuations are likely to be non local (e.g. filaments erupting from the separatrix) and the background is an approximate equilibrium only for the local instabilities that have a shorter timescale than the background variations.       

\subsection{General model} \label{sec1.1}

Following \cite{Militello2016c}, we assume that individual filaments have a well defined shape, $\Lambda(x,w)$, in the radial direction, $x$, and that ther are parametrised by, $w$, that measures the perpendicular width of the filament. We set the separatrix at $x=0$ with the SOL located at $x>0$. The evolution of the filament can be represented as:
\begin{equation}
\label{2} 
\eta_i(x,t)=\eta_{0,i}F_i(t)\Lambda\left(x-\int_0^t V_i(t')dt',w_i\right) 
\end{equation}
where $t$ is time, $\eta_0$ represents the initial amplitude of the thermodynamic variable associated with the filament (i.e. density or temperature), $F(t)$ describes the reduction of this amplitude due to generic parallel losses, $V(t)$ is the radial velocity of a filament and the subscript $i$ labels different filaments. 

We assume that filaments do not interact with each other while travelling in the SOL even if new filaments move in the wake of older filaments and add their amplitude to generate the time signals. Therefore, at each time $t$ and position $x$ the train of filaments would produce a signal given by:
\begin{equation}
\label{3}
\theta(x,t) = \sum_{i=1}^\infty \eta_i\left(x,t-t_i\right),
\end{equation}
where $t_i$ is the time at which the \textit{i}th filament crosses the separatrix. At each radial position, $x$, the signal produced by Eq.\ref{3} is also known as shot noise and it has the features of a Poisson process \cite{Pecseli2000,Garcia2012}, which is an ergodic process. 

To go from the filaments' motion to the radial profile of the thermodynamic variable, $\Theta(x)$, is to time average $\theta(x,t)$:
\begin{equation}
\label{4}
\Theta(x) =   \overline{\sum_{i=1}^{\infty}\eta_i(x,t-t_i)}
\end{equation}
where we have defined $\overline{\cdots}\equiv \displaystyle{\lim_{\Delta T \rightarrow \infty}}\Delta T^{-1}\int_{0}^{\Delta T}\cdots dt$. 

It is useful to start by calculating $\Theta(x)$ in a finite interval $\Delta T$, in which $K$ filaments will contribute to the signal. We therefore have that $\Theta_{\Delta T}(x) = \frac{1}{\Delta T}\int_0^{\Delta T}dt \sum_{i=0}^K \eta_i(x,t-t_i)$. In general, this is an ill-defined quantity, since $K$ is a statistical variable that depends on $\Delta T$. Indeed, the number of filaments in a given interval is assumed to behave according to a Poisson distribution, $P_K=\lambda^K e^{-\lambda}/K!$, $\lambda=\Delta T/\tau_w$ and $\tau_w$ is the average time between filaments, i.e. their waiting time. Using ergodicity, we replace the time average with an ensemble average over the possible statistical outcomes. This leads to $\Theta_{\Delta T}(x) = <\theta(x,t)>$ where he operator $<\cdots>= \int_0^{\infty}P_{\eta_0}d\eta_{0,i}\int_0^{\infty}P_w dw_i\sum_{K=0}^\infty P_K\int_0^{\Delta T} dt_i P_{t_i}\cdots$ represents an ensemble average over the waiting times. Here, $P_{t_i}=1/\Delta T$ is the homogeneous probability distribution of the arrival times associated with a Poisson process, while $P_{\eta_0}$ and $P_w$ represent the probability distribution functions of the initial amplitudes and widths of the filaments. It can be shown \cite{Militello2016c} that:
\begin{equation}
\label{4b}
\Theta(x) = \{\eta(x,t)\} = \frac{1}{\tau_w} \int_{-\infty}^{\infty}dt\int_0^{\infty}d\eta_0 \int_0^{\infty}dw \left[\eta(x,t) P_{\eta_0}(\eta_0)P_w(w)\right],
\end{equation}
where we have defined the ensemble average operator with curly brackets. Note that the order in which the integrals are performed is not relevant since $\eta_0$ and $w$ do not depend on time or on each other (the latter statement is an assumption). 

To compare with experimental observations, it is useful to determine other properties of the signals, such as successive statistical moments. These, of course, characterise more accurately the time series and provide a more stringent constraint for the model validation. After the time average given in Eq.\ref{4b}, it is natural to study the variance of the signal, which broadly speaking represents the amplitude of the fluctuations. This is defined as:
\begin{equation}
\label{4c}
\sigma^2(x) = \overline{\left[\sum_{i=0}^{\infty} \eta_i(x,t-t_i)-\Theta(x)\right]^2}=\overline{\left[\sum_{i=0}^{\infty} \eta_i(x,t-t_i)\right]^2}-\Theta(x)^2.
\end{equation}
Applying the same statistical procedure described above to Eq.\ref{4c}, we obtain the variance skewness and kurtosis:
\begin{eqnarray}
\label{4d}
\sigma^2(x) &=& \{\eta(x,t)^2\}, \\
\label{4e}
S(x) &=& \frac{ \{\eta(x,t)^3\}  }{ \{\eta(x,t)^2 \}^{3/2} }, \\
\label{4f}
K(x) &=& \frac{\{\eta(x,t)^4 \}}{\{\eta(x,t)^2 \}^{2}}.
\end{eqnarray}

In the next Sections, we discuss the form of $F(t)$ and $V(t)$ which we justify on the basis of single filament physics. Crucially, several interesting properties of the profiles emerge naturally from a reasonable choice for these functions.   

\section{Trajectory of the filaments} \label{sec2}

We now consider a single filament with given initial amplitude and size and discuss what determines its trajectory in the SOL. Within our model it is equivalent to assign specific functional dependencies to $V(t)$ and $F(t)$. The form of these functions is justified in the following subsections, while Section \ref{sec3.0} discusses how these assumptions can translate into specific models that can be applied to the statistical framework. 

\subsection{Parallel dynamics: model and equations} \label{sec2.1}

Our model takes a Lagrangian perspective on the dynamics of the filaments. We assume that the filament maintains its coherence while translating in the perpendicular direction. In this case, the normalised equations governing the parallel decay of the filament are given by \cite{Braginskii1965,Havlickova2013}:
\begin{eqnarray}
\label{5}
\frac{\partial n}{\partial t}   &=& -\frac{\partial}{\partial s} (nv) +S_n,\\
\label{6}
\frac{\partial (nv)}{\partial t}  &=& -\frac{\partial}{\partial s} \left[nv^2+p_e+p_i\right] +S_v, \\
\label{7}
\frac{3}{2}\frac{\partial p_e}{\partial t} &=& -\frac{\partial}{\partial s} \left(\frac{5}{2}p_ev+\frac{\epsilon_e}{\Lambda_c}\frac{2}{7}\frac{\partial T_e^{7/2}}{\partial s}\right)+v\frac{\partial p_e}{\partial s} +3\Lambda_c \frac{n^2}{T_e^{3/2}}(T_i-T_e)+S_{p,e}, \\
\label{8}
\frac{3}{2}\frac{\partial p_i}{\partial t} &=& -\frac{\partial}{\partial s} \left(\frac{5}{2}p_iv+\frac{\epsilon_i}{\Lambda_c}\frac{2}{7}\frac{\partial T_i^{7/2}}{\partial s}\right)+v\frac{\partial p_i}{\partial s} -3\Lambda_c \frac{n^2}{T_e^{3/2}}(T_i-T_e)+S_{p,i},
\end{eqnarray}
where $n$, $v$, $p_e$ and $p_i$ are the evolved variables, which represent plasma density, parallel velocity, electron and ion pressure (normalised to characteristic values unless otherwise stated). The ion and electron temperature is defined through $T_s \equiv p_s/n$, as usual. The parallel direction is measured by the coordinate $s$. The normalisation is based on a typical length and timescales given by the midplane to target connection length, $L$, and the transit time, $L/c_s$, where the ion sound speed is $c_s \equiv \sqrt{(T_e+T_i)/m_i}$, with temperatures measured here in eV. Also, we define an electron to ion mass ratio corrected colisionality $\Lambda_c\equiv (m_e/m_i)^{1/2}L_\parallel/\lambda_{ei}$, where $\lambda_{ei}$ is the mean free path calculated with the characteristic $T_{e,i}=T_{e,i,0}$ and $n=n_0$, $\epsilon_e \approx 3.2$ and $\epsilon_i\equiv 3.9 (\tau_i/\tau_e)(m_e/m_i)A^{1/2}\approx 0.09A^{1/2}$ for an ion charge $Z=1$ and an ion to proton mass ratio $A$, where $\tau_{i,e}$ are the collision times for ions and electrons. We have assumed negligible electron inertia and no parallel current (i.e. ambipolar behaviour, so that both species move at the same velocity). The system is closed by collisional parallel conductivity, represented by the second term on the right hand side of Eqs.\ref{13} and \ref{14}, and by collisional heat exchange between species, which in the normalised form becomes $3\Lambda_c n^2(T_i-T_e)/T_e^{3/2}$. 

The terms $S_n$, $S_v$ and $S_p$ represent sources/sinks of particles, momentum and pressure. It is useful to estimate $S_n\approx n/\tau_{iz}$ and $S_v\approx -nv/\tau_{CX}$, the former mainly given by ionisation and the latter by charge exchange, assuming neutrals at rest and with a constant neutral density. More specifically, $\tau_{iz}^{-1} = (L/c_s) n_n\sigma_{iz}$ and $\tau_{CX}^{-1} = (L/c_s) n_n\sigma_{CX}$  ($n_n$ and $\sigma$ are the dimensional neutral density and relevant interaction rate coefficient). Here we assumed that recombination can be neglected, but this approximation might be invalid close to detachment. In addition, ionization is a sink for the electron pressure, $S_{p,e}\approx -nT_{iz}/\tau_{iz}$, where $T_{iz}$ is the ionisation energy expressed in eV, while charge exchange also removes ion pressure, $S_{p,i}\approx -(3/2)nT_i/\tau_{CX}+(1/2)v^2/\tau_{CX}+(1/2)v^2/\tau_{iz}$, assuming that the energy of the neutrals is much smaller than the energy of the ions. Note that we have neglected also the cooling rate due to excitation, but its effect could be mocked up by artificially increasing the ionization energy \cite{Togo2013}.

The boundary conditions for Eqs.\ref{5}-\ref{8} must model the presence of the Debye sheath at the target. Hence, at the end of the domain, we assume that the velocity becomes sonic, $V=\sqrt{T_e+T_i}$. In addition, the energy flux through the sheath is set at $5p_e\sqrt{T_e+T_i}$ for the electrons and at $3.5p_i\sqrt{T_e+T_i}$ for the ions. The target density is free to evolve, otherwise the system would be overdetermined \cite{Easy2014,Havlickova2013}. We also assume an up/down symmetric configuration, so that a symmetry plane exists in the middle of the domain, in which $V=\partial_s n = \partial_s p_e = \partial_s p_i = 0$.

\subsection{Parallel dynamics: double timescale} \label{sec2.2}

We start by simplifying the system to elucidate some of its features. This can be done, for example, in the limit of small collisionality, $\Lambda_c \ll 1$, where the density and electron temperature dynamics decouple, the latter being much faster because it is driven by the efficient parallel heat conduction \cite{Fundamenski2007}. From Eq.\ref{6}, assuming no sources, we find that parallel pressure variations are balanced by velocities of the order of the sound speed, $v\sim \sqrt{T_e+T_i}$. This means that a ballooned filament with parallel length scale $L_\parallel<L$ drains its density following $\partial_t n \sim -\sqrt{T_e+T_i}(L/L_\parallel)n$. If we assume a roughly constant temperature (the square root dependence is anyway weak), this gives an exponential decay in time with a timescale $\tau_n\sim L_\parallel/L$, i.e. shorter than the transit time (which in our normalisation this equals 1). 

The temperature can be removed towards the divertor by two mechanisms, conduction, $\partial_t T_s \sim -T_s^{7/2}n^{-1}(\epsilon_s/\Lambda_c)(L/L_\parallel)^2$, or convection, $\partial_t T_s \sim -T_s^{3/2}\sqrt{(1+\Theta_s)}(L/L_\parallel)$ (assuming no sources/sinks and weak coupling), where $\Theta_i=\Theta_e^{-1}\equiv T_e/T_i$ is assumed constant. The dominance of one mechanism over the other depends on the collisionality in the SOL, as discussed, for example, in \cite{Fundamenski2007}. In any case, both limits lead to solutions that decay with a timescale that changes with $t$ due to the nonlinearity in $T_s$ of the heat conductivity and the sound speed. The fast conductive timescale becomes longer as the collisionality increases and eventually becomes comparable to the convective \cite{Fundamenski2007}. In other words, as temperature decreases, so does the efficiency of its exhaust mechanism. An important limitation of the calculation above is that we assumed a constant length scale for the filament. However, in the presence of ballooned filaments, the convective and conductive terms tend to remove the parallel inhomogeneities, thus quickly increasing $L_\parallel/L$. This, in turn, has the effect of making the exhaust timescales longer by a factor $(L_\parallel/L_{\parallel,0})^\alpha$, where $L_{\parallel_0}$ is the initial value determined by core turbulence and $\alpha=1$ for convection and $\alpha=2$ for conduction.  

When the parallel gradients are removed, the relevant timescale of the system is set by the boundary conditions, and in particular, the sheath physics. The \textit{dimensional} characteristic time for density removal by the sheath is estimated by $\tau_{n,sh}=N/\partial_tN|_{sh}\sim n\mathcal{A}L/(nc_sA) \approx L/c_s$, where $\mathcal{A}$ is the perpendicular area of the flux tube, $N$ the total density in the flux tube. Upon applying the normalisation, this time becomes equal to unity. When $L\sim L_\parallel$, homogenisation and sheath timescales are roughly comparable, due to the fact that they are both regulated by ion inertia. On the other hand, ballooned filaments can easily give an homogenisation timescale 5-6 times faster than the one associated with the exhaust at the target since $L_\parallel$ is a fraction of $L$. 

At the sheath, the particle flux also determines, to a large extent, the energy flux, so that $\tau_{T_s,sh}$ are of order unity. The previous estimate is subject to a few caveats as the sheath transmission coefficient for the electrons would shorten the electron energy timescale by a factor 5, while the ion energy timescale is much more difficult to assess, due to the fact that their kinetic energy is not negligible.     

If the homogenisation and the sheath removal timescales are sufficiently separated, the time evolution of the thermodynamic quantity undergoes a transition and shows a double feature. This is important, since the presence of two timescales in the parallel dynamics of the filaments naturally leads to the presence of two length scales in the radial profiles, as it is shown in Sec.\ref{sec3.2}. For the density, ballooning is needed to have a double feature as $\tau_n/\tau_{n,sh} = L_\parallel/L$. The timescale separation automatically occurs for electron temperature, since $\tau_{T_e}/\tau_{T_{e,sh}}\sim (L_\parallel/L)^2(\Lambda_c/\epsilon_e)$, as the numerator is determined by electron physics while the denominator by ion physics.     

The results presented above are used in Section \ref{sec3.2}, where we explore the consequences of having multiple timescales on SOL profiles.

\subsection{Parallel dynamics: effect of the neutrals} \label{sec2.3}

It is interesting to determine how sources and sinks due to neutrals can affect the parallel evolution of the filament. Under the assumption that the neutrals are stationary and at constant density, charge exchange plays a dominant role, due to the fact that it has a larger cross section than ionisation (or recombination). In Eq.\ref{6} the balance of the pressure drive shifts from the advective terms, $v\sim \sqrt{T}$, to the charge exchange term, $nv\sim \tau_{CX} \partial n/\partial s$ when $\tau_{CX}$ becomes smaller than the parallel transit time (i.e. when the neutral density increases). 

Replacing this estimates in Eq.\ref{5} shows that the advective nature of the parallel transport turns into conductive when neutrals are sufficiently dense. This implies that the new dimensional timescale associated with the density removal is ordered as $(L/c_s)^2/\tau_{CX}$, which becomes longer as $n_n$ increases. In other words, the charge exchange induced friction with the neutrals slows down the plasma motion towards the target, hence "clogging" its exhaust. Ionisation, can further increase the upstream density by injecting new plasma. 

Importantly, in the collisionality regimes where thermal conduction dominates over advection, the electron temperature dynamics should not be significantly modified by charge exchange, thus leaving the profiles unaltered. Ionisation, on the other hand, affects the dynamics through the sink terms as the electron temperature decreases faster in the presence of a significant neutral population. The experimental observation that electron temperature profiles seem to be unaffected in the presence of density broadening seems to suggest that ionisation should not play a dominant role in the phenomenon. 

The results in the Subsection are applicable transversally in all the models discussed in Section \ref{sec3.0}. If the neutral density has a radial profile (larger at the walls, where recycling occurs, smaller at the separatrix) this could induce again multiple timescales in the problem. In general, however, we see charge exchange as a mechanism that increases the timescale globally as fuelling increases and hence more related to the broadening mechanism (see Section \ref{sec3.1}). 

\subsection{Perpendicular filament velocity} \label{sec2.4}

Theoretical scaling laws and numerical simulations of 2D and 3D isolated filaments \cite{Krasheninnikov2001, Garcia2007,Omotani2015} suggest that the radial velocity of the filament is determined by its perpendicular size and amplitude:
\begin{equation}
\label{9}
\begin{cases}
V = v_{in} T_0^{1/2}w^{1/2}A^{1/2} & w \ll w_{cr} \\ 
V = v_{sh} T_0^{3/2}w^{-2}\left(\frac{A}{1+0.3 A}\right)  & w \gg w_{cr},
\end{cases}
\end{equation}
where $A$ is a measure of the fluctuation amplitude defined as the maximum of $(p-p_0)/p_0$ with $p$ and $p_0$ the pressure of the filament and of the background in which it is moving (created by the wakes of the previous perturbations or leakage from the core). In addition, $T_0$ is the background temperature, $w_{cr}$ is a critical filament size that separates the inertial from the sheath regime \cite{Garcia2006,Omotani2015}, $v_{in}$ and $v_{sh}$ are constant prefactors relevant for the inertial and sheath regime respectively. 

Experimental results \cite{DIppolito2011,Theiler2009} show that these scaling laws produce an upper bound to the velocities. This might be due to the different ratio of density and temperature perturbations for the same pressure, which would produce different radial velocities \cite{Militello2016b,Walkden2016b}. Also, from experiments it appears that the radial velocity does not have a large variance. This might be due to the fact that the filaments usually sit close to the plateau between the sheath and the inertial regimes, where width variations have little effect on the filaments \cite{Theiler2009,Kirk2016,Militello2016b}.  

Clear experimental measurements of $A$ are not available as estimating the background profiles is a difficult task. However, Boedo \textit{et al.} \cite{Boedo2003} suggest that $A$ remains roughly constant in different experimental conditions and at all radii. As a consequence, we assume that it is independent from time and other parameters, but we do retain its effect on the radial velocity definition as it determines faster motion for filaments with larger amplitude above the background, an effect observed also in \cite{Zweben2016}. A simplified but reasonable form for the velocity is therefore:
\begin{equation}
\label{10}
V \approx v_0A^{\alpha}\frac{\sqrt{w/w_{cr}}}{1+(w/w_{cr})^{5/2}}
\end{equation}   
where $v_0$ contains the temperature dependence and a Pade' approximation is used to capture the transition between the two filament regimes. This was confirmed by 3D simulations of isolated filaments, which showed such a transition \cite{Easy2014,Easy2016}. In the following we will consider large filaments, for which we can take $A\propto \eta_0/\Theta(0)$. 

The results presented here will be used in Sections \ref{sec3.4} and \ref{sec3.5}, where an amplitude and width dependent velocity combines with the statistics of the filaments in order to generate non-exponential profiles.

\section{Filamentary models and profile generation} \label{sec3.0}

In order to understand the mechanism behind the profile formation, we compare the predictions of different models with the experimentally observed features of the flattening and the broadening. In particular, acceptable models must be able to explain: I) the change in the decay length observed between near and far SOL, i.e. the flattening; II) the increase of the relative fluctuation level in the far SOL. When possible we derive exact analytic expressions for the profiles, otherwise we employ numerical integration to obtain trends. All the analytic and semi-analytic calculations are verified using synthetic signals generated numerically using trains of filaments with consistent statistical assumptions. 

\subsection{Constant velocity, exponential decay} \label{sec3.1}

We start with the simplest case, corresponding to $P_{\eta_0}(\eta_0)=\delta(\eta_0-\eta_*)$ and $P_w(w)=\delta(w-w_*)$ where $\delta$ is the Dirac delta function. Assuming $\alpha=0$, no $w$ dependence in $V$, $F(t)=e^{-t/\tau}$ and $\Lambda(x,w) = e^{x/w}H(-x)$, where $H(x)$ is the Heaviside function, we have:
\begin{equation}
\label{14b}
\Theta(x) = \frac{\eta_*}{\frac{\tau_w}{\tau}\left(1+\frac{V\tau}{w_*}\right)}e^{-\frac{x}{V\tau}},
\end{equation} 
and:
\begin{equation}
\label{14c}
\frac{\sigma(x)}{\Theta(x)} = \frac{\sqrt{2}}{2}\sqrt{\frac{\tau_w}{\tau}\left(1+\frac{V\tau}{w_*}\right)}.
\end{equation} 
The details of the analytic calculations are given in the Appendix. The profile given by Eq.\ref{14b} has a constant decay length given by $L_\Theta=V\tau$, which suggests that the flattening cannot occur within this model. In addition, the relative fluctuation amplitude (proportional to $\sqrt{\eta_*/\Theta(0)}$) and the other statistical moments of the signal do not vary with radial position, in disagreement with the experimental evidences. 

\begin{figure}
\includegraphics[height=8cm,width=12cm, angle=0]{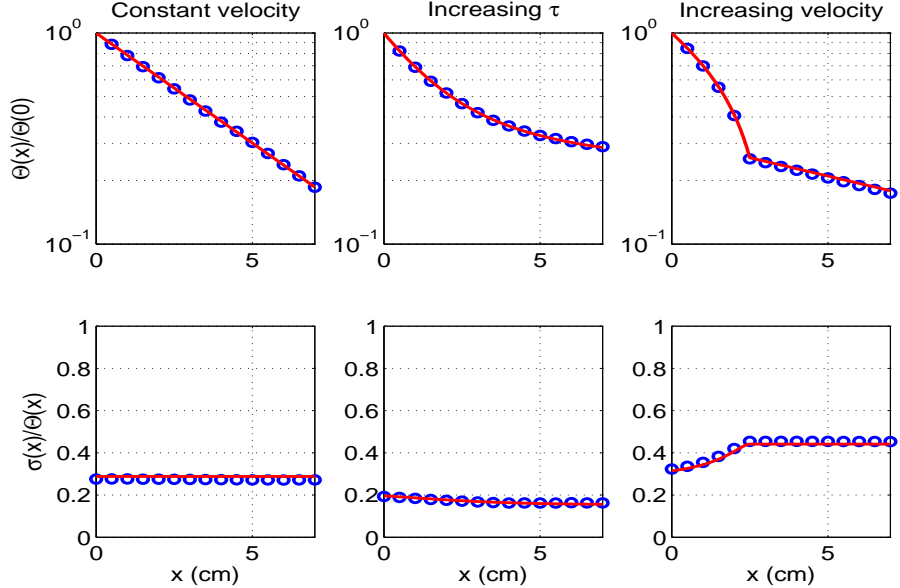}
\caption{Comparison between the mean radial profile (top row) and normalised fluctuation amplitude (bottom row) between the analytic predictions (solid lines) and the the simulated signals (circles). Equations \ref{14b} and \ref{14c} are used for the first column, Eqs.\ref{14d} and \ref{14e} for the second and Eqs. \ref{14g} and \ref{14i} for the third.}
\label{fig2}
\end{figure}

A comparison between the analytic predictions, Eqs\ref{14b}-\ref{14c}, and the time average of a synthetic signal is given in the first column of Fig.\ref{fig2}. For the synthetic signal we generated a random train of filaments as a Poisson process (for a total of 1200 filaments). We focused on the density field (i.e. $\Theta(x)$ stands for the density profile), we used the same $\Lambda(x,w)$ of the model, $\eta_*=0.1\times 10^{13} cm^{-3}$, $\tau_w/\tau=0.05$, $w_*=2 cm$ and $V\tau= 4cm$, which are representative of MAST L-mode \cite{Militello2016,Kirk2016}.    

It is useful to notice that increasing the velocity alone broadens the profiles, but also shifts the separatrix value downwards (for the density, this is not observed in experiments). For the parameters discussed above, doubling the velocity doubles the decay length (from $4 cm$ to $8 cm$), but it also reduces $\Theta(0)$ by $40\%$, from $0.66 \times 10^{13} cm^{-3}$ to $0.4 \times 10^{13} cm^{-3}$. An increase in $\tau$ leads to broadening as well, but is accompanied with a larger $\Theta(0)$. In our example, doubling $\tau$ leads to $\Theta(0)= 0.8 \times 10^{-13} cm^{-3}$. Note that the amplitude of the filaments and their perpendicular size only affect the absolute value of the profile, but not its shape. 

The constant velocity model captures the naive approach used in several empirical interpretations of the profile broadening: a transition in the filament dynamics that produces larger perpendicular velocities leads to broader profiles. Our analysis shows that this approach is an oversimplification which cannot explain relevant experimental observations associated with the flattening of the profiles (the decay length is constant) and with the radially changing statistics (which remain the same). 

\subsection{Constant velocity, double exponential decay} \label{sec3.2}

In Section \ref{sec2.2} we discussed the possibility that filaments decay with two distinct timescales. Taking $F(t) = \alpha_1e^{-t/\tau_1}+\alpha_2e^{-t/\tau_2}$, where $\tau_1<\tau_2$ and $\alpha_1+\alpha_2=1$, Eq.\ref{14b} can be trivially extended to describe profiles with a double exponential nature. 

In particular, for the profile we obtain:
\begin{equation}
\label{14d}
\Theta(x)= \alpha_1\frac{\eta_*}{\frac{\tau_w}{\tau_1}\left(1+\frac{V\tau_1}{w_*}\right)}e^{-\frac{x}{V\tau_1}}+\alpha_2\frac{\eta_*}{\frac{\tau_w}{\tau_2}\left(1+\frac{V\tau_2}{w_*}\right)}e^{-\frac{x}{V\tau_2}},
\end{equation}
while the fluctuations follow:
\begin{equation}
\label{14e}
\frac{\sigma(x)}{\Theta(x)} = \frac{\sqrt{2}}{2}\sqrt{\frac{\tau_w}{\tau_1}\left(1+\frac{V\tau_1}{w_*}\right)}G(x),
\end{equation}
where $G(x)$ represents the variation due to the presence of the second exponential decay and is given by:
\begin{equation}
\label{14f}
G(x) = \frac{\left[ 1 + 2\frac{\alpha_2}{\alpha_1}\frac{1+\frac{w}{V\tau_1}}{1+\frac{1}{2}(\frac{w}{V\tau_1}+\frac{w}{V\tau_2})}e^{-\frac{x}{w}\left(\frac{w}{V\tau_2}-\frac{w}{V\tau_1}\right)} + \left(\frac{\alpha_2}{\alpha_1}\right)^2\frac{1+\frac{w}{V\tau_1}}{1+\frac{w}{V\tau_2}}e^{-2\frac{x}{w}\left(\frac{w}{V\tau_2}-\frac{w}{V\tau_1}\right)} \right]^{1/2}}{1 +\frac{\alpha_2}{\alpha_1}\frac{1+\frac{w}{V\tau_1}}{1+\frac{w}{V\tau_2}}e^{-\frac{x}{w}\left(\frac{w}{V\tau_2}-\frac{w}{V\tau_1}\right)} }.
\end{equation}
In the second column of Fig.\ref{fig2} these results are compared to a synthetic signal generated as discussed in the previous Subsection, with $\eta_* = 0.1e13 cm^{-3}$, $\tau_w/\tau_1=0.05$, $w_*=2 cm$ and $V\tau_1= 1.6cm$, $\tau_2/\tau_1=20$, $\alpha_1=0.8$.    

Importantly, $G(x)$ always decreases with $x$, so that the fluctuation level described by this model behaves in the opposite way with respect to the experimental observations. On the other hand, the reduction of $G(x)$ is quite weak ($<40\%$) for reasonable plasma parameters and ranges of $x$. However, this shows that parallel dynamics alone is not sufficient to provide a satisfactory explanation for the flattening consistent with all measured data. 

\subsection{Time or space dependent velocity, exponential decay} \label{sec3.3}

It is interesting to evaluate what happens to the profiles if the velocity changes in time or in space. From a theoretical perspective, a filament might accelerate because of the reduced background through which it moves \cite{Omotani2015} or because of the increased resistivity in the far SOL which reduces the sheath dissipation \cite{Easy2016}. Starting from the approximations of Section \ref{sec3.1}, we further introduce the simplest time dependent velocity: $V=V_0+(V_1-V_0)H(t-t_0)$, which leads to: %$X(t) = V_0t+(V_1-V_0)(t-t_0)H(t-t_0)$. This implies: $X^{-1}(x) = \frac{x}{V_0}+(x-V_0t_0)\left(\frac{1}{V_1}-\frac{1}{V_0}\right)H(x-V_0t_0)$. Hence:
\begin{equation}
\label{14g}
\begin{cases}
\Theta(x) = \frac{\eta_*}{\frac{\tau_w}{\tau}\left(1+\frac{V_0\tau}{w_*}\right)}e^{-\frac{x}{V_0\tau}}\left[1-\frac{\frac{V_1\tau}{w_*}-\frac{V_0\tau}{w_*}}{1+\frac{V_1\tau}{w_*}}e^{\left(1+\frac{V_0\tau}{w_*}\right)\left(\frac{x}{V_0\tau}-\frac{x_0}{V_0\tau}\right)}\right] & x<x_0 \\
\Theta(x) =  \frac{\eta_*}{\frac{\tau_w}{\tau}\left(1+\frac{V_1\tau}{w_*}\right)}e^{-\frac{x}{\tau V_1}+\frac{x_0}{\tau V_1}-\frac{x_0}{\tau V_0}} &x>x_0
\end{cases}
\end{equation}
with $x_0\equiv V_0t_0$. Note that this formalism and the equations above can be straightforwardly applied to a spatially dependent change, if one interprets $x_0$ as a fixed point independent from $V_0$. If $V_1>V_0$, it can be easily shown that the typical lengths scale of the profiles is, for $x<V_0t_0$, a fraction of $\tau V_0$ that decreases with $x$, while for $x>V_0t_0$ it is constant and equal to $\tau V_1$. On the other hand, a filament slowing down would produce the opposite behaviour, with flatter profiles close to the separatrix. 

The amplitude of the fluctuations is given by:
\begin{equation}
\label{14i}
\begin{cases}
\frac{\sigma(x)}{\Theta(x)} = \frac{\sqrt{2}}{2} \sqrt{\frac{\tau_w}{\tau}\left(1+\frac{V_0\tau}{w}\right)}\frac{\left[1-\frac{\frac{V_1\tau}{w_*}-\frac{V_0\tau}{w_*}}{1+\frac{V_1\tau}{w_*}}e^{2\left(1+\frac{V_0\tau}{w_*}\right)\left(\frac{x}{V_0\tau}-\frac{x_0}{V_0\tau}\right)}\right]^{1/2}}{1-\frac{\frac{V_1\tau}{w_*}-\frac{V_0\tau}{w_*}}{1+\frac{V_1\tau}{w_*}}e^{\left(1+\frac{V_0\tau}{w_*}\right)\left(\frac{x}{V_0\tau}-\frac{x_0}{V_0\tau}\right)}} & x<x_0 \\
\frac{\sigma(x)}{\Theta(x)} = \frac{\sqrt{2}}{2} \sqrt{\frac{\tau_w}{\tau}\left(1+\frac{V_1\tau}{w}\right)} & x>x_0
\end{cases}
\end{equation}
Equations \ref{14i} describe a continuous function that is growing in the range $0<x<x_0$ if $V_1>V_0$ (it decreases for the opposite condition).

The third column of Fig.\ref{fig2} shows the mean profile and the relative fluctuation amplitude as a function for this model ad compares it with a synthetic signal obtained using the same parameters used in Section \ref{sec3.1} with $V_1/V_0=3$ and $x_0=2.5 cm$.

\subsection{Statistically distributed amplitudes and amplitude dependent velocity} \label{sec3.4}

The results in the previous three Subsections assumed filaments with a single width and amplitude. We now introduce a distribution of amplitudes that, based on experimental observations \cite{Garcia2013}, we chose to be exponential: $P_ {\eta_0}=\eta_*^{-1}e^{-\eta_0/\eta_*}$. Extension of the results in Sections \ref{sec3.1}-\ref{sec3.3} to distributed amplitudes is, in principle, straightforward as it is sufficient to multiply the mean and the variance by the PDF of $\eta_0$ and integrate (note that this is not true for the relative fluctuation amplitude). By doing this, it is easy to check that Eq.\ref{14b} would remain the same, while Eq.\ref{14c} would not have the $\sqrt{2}/2$ factor on the right hand side. 

When also an amplitude dependent velocity is introduced in the problem (i.e. $\alpha\neq 0$), a closed solution cannot be found in general and we therefore used numerical tools to understand this effect. We can, however, have a qualitative understanding of the expected features of the profiles using the results in Section \ref{sec3.1} by replacing $V=const$ with $V(\eta_0) \propto \eta_0^\alpha$ and interpreting the final profile as a weighted sum of single amplitude profiles. If $\alpha>0$, it is easy to see that filaments with larger (smaller) amplitude give flatter (steeper) contributions to the profiles and are also associated with larger (smaller) relative fluctuations. This implies that $\Theta(x)$ and $\sigma(x)/\Theta(x)$ change in space, in particular they both increase at larger radii. We conclude that an amplitude dependent velocity always produces a flattening of the profiles. By extension, the flattening produced by an increasing exponential decay, Eq.\ref{14d}, or velocity, Eq.\ref{14g}, is enhanced in the presence of distributed amplitude dependent velocities.

The flattening can be quantified in a practical (but not general) way by introducing the parameter $f\equiv L_{\Theta,w}/L_{\Theta,s}$, where $L_\Theta\equiv -\Theta(x)/\Theta'(x)$ is the length scale associated with the mean profile and the subscripts $s$ and $w$ are the values at the separatrix and at the first wall. When this is done for the MAST parameters discussed in Section \ref{sec3.1}, we find that the flattening is more evident at larger values of $\alpha$ and lower of $\eta_*$. The latter result depends on the choice of $\Lambda(x,w)$ as more symmetric shapes (e.g. Gaussian) do not lead to an increase of $f$ at small amplitudes. 

\begin{figure}
\includegraphics[height=8cm,width=12cm, angle=0]{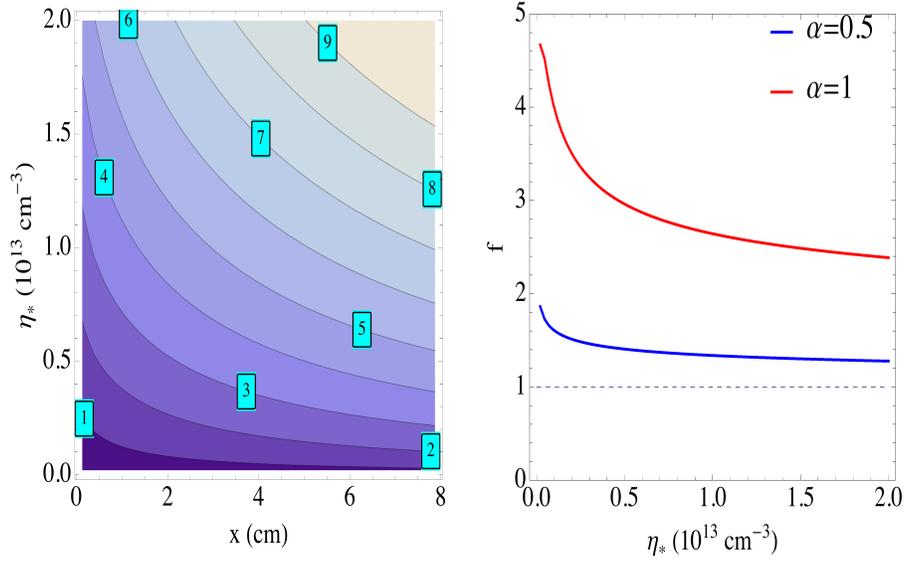}
\caption{Left: contour plot of $L_\Theta$ for $\alpha=1$ as a function of the radial position, $x$ and of the mean initial filament amplitude, $\eta_*$, showing the increase of the decay length in the far SOL. Each contour line is marked with the corresponding value of $L_\Theta$ in cm. Right: flattening parameter, $f$, as a function of the mean initial filament amplitude. The two solid curves are associated with different values of the exponent $\alpha$. The curves are plotted for the same parameters used in Section \ref{sec3.1}.}
\label{fig3}
\end{figure}
\begin{figure}
\includegraphics[height=7cm,width=14cm, angle=0]{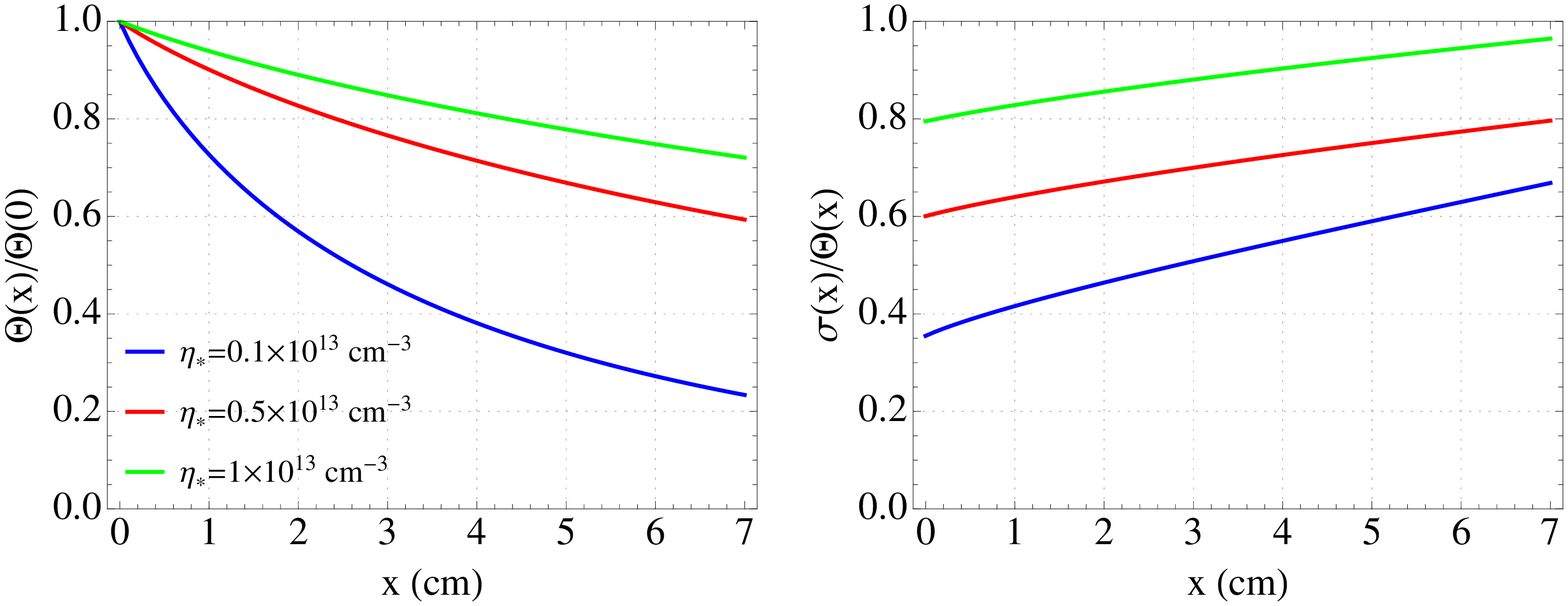}
\caption{Left: mean profiles normalised to the separatrix value for different mean initial amplitude. Right: corresponding normalised fluctuation amplitude. The curves are plotted for the same parameters used in Section \ref{sec3.1} and $\alpha=1$.}
\label{fig4}
\end{figure}

Figure \ref{fig3} shows $f$ as a function of of the mean amplitude and of $\alpha$ and reveals that a variation of the former does not change significantly the level of flattening. Fig.\ref{fig4} shows that the relative fluctuation amplitude increases as a function of radius when amplitude dependent velocities with a statistical distribution are taken into account. It is interesting to notice that a larger $\eta_*$ coincides with profiles that seem to generate a broadening as the decay length increases more rapidly in the far than in the near SOL. The increase in the mean amplitude also leads to a higher $\Theta(0)$, which is compatible with experimental observations. This result is obtained assuming that an increased fuelling leads to a higher $\eta_*$ without affecting $\tau_w$ and implies an increased particle flux. %As a matter of fact, it can be proven that $\sigma(x)/\Theta(x)$ is a monotonically growing function of $x/[V(\eta_*)\tau]$, that its steepness increases as a function of $V(\eta_*)\tau/w$ and that its absolute value depend on $\sqrt{\tau_w/\tau}$.

\subsection{Statistically distributed width and width dependent velocity} \label{sec3.5}

As for the amplitude, it is straightforward to include in the calculation a distribution of widths. Unfortunately, even for the simplest cases, an analytic solution is not available, but numerical calculations are not demanding. To our knowledge, a detailed discussion of the functional form of the distribution of filaments' widths in the SOL is not available in literature. On the other hand, \cite{BenAyed2009} and \cite{Kirk2016} provide quantitative descriptions of the probability distribution function in MAST. For our purposes, we employ a log-normal distribution, which is in qualitative agreement with these references: $P_w = (wS\sqrt{2\pi})^{-1}e^{-(\log w -\mu)^2/2S^2}$, with $\mu=\log(w_*/\sqrt{1+\sigma_w^2/w_*^2})$, $S^2=\log(1+\sigma_w^2/w_*^2)$, where $w_*$ and $\sigma_w^2$ are the mean and the variance of the distribution. 

\begin{figure}
\includegraphics[height=8cm,width=12cm, angle=0]{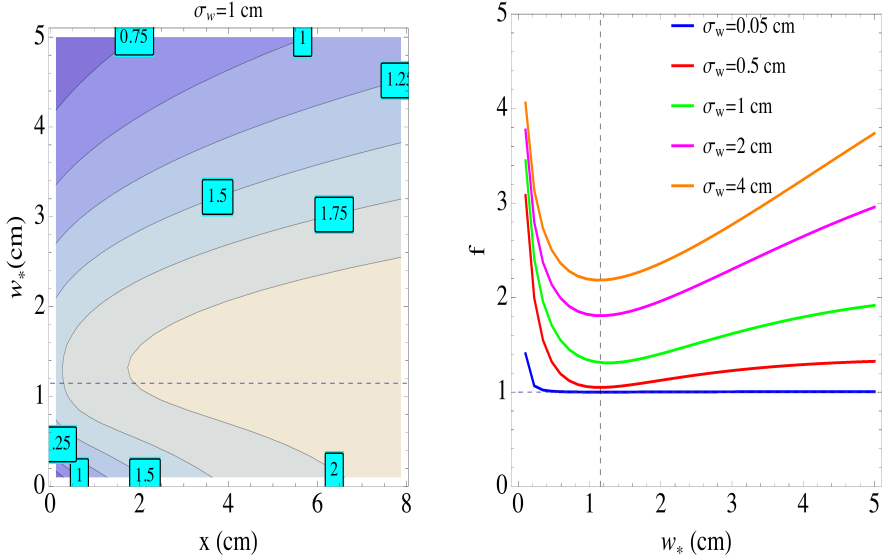}
\caption{Left: contour plot of $L_\Theta$ for $\sigma_w=1$ as a function of the radial position, $x$ and of the mean filament width $w_*$ showing the increase of the decay length in the far SOL. Each contour line is marked with the corresponding value of $L_\Theta$ in cm. Right: flattening parameter $f$ as a function of the mean filament width. The solid curves are associated with different values of the variance $\sigma_w$. The curves are plotted for the same parameters used in Section \ref{sec3.1} and for a $w_{cr}=2 cm$. The position of $w_m=2^{1/5} cm$ is plotted in both sub-plots with a dashed line.}
\label{fig5}
\end{figure}

If no width dependence is included in the velocity, numerical solutions show that the results presented in Sec.\ref{sec3.1} remain substantially unchanged. This is easily understood noticing that the radial dependence in both the average profile and in the variance filters out of the integrals. On the other hand, when Eq.\ref{10} with $\alpha=0$ is used to describe the velocity, the profiles show a clear flattening, the magnitude of which depends on the mean filament width and on the variance of the distribution. In addition, the relative fluctuation amplitudes grows in the far SOL. We find that a larger variance corresponds to larger $f$, as shown in Fig.\ref{fig5}, which is understandable as the range of filament velocities sampled increases. Also, the flattening is more evident at low and high $w*$ and has a minimum at an intermediate width $w_m$, see Fig.5. The width $w_m$ corresponds to the maximum of the velocity curve, Eq.\ref{10}. The flattening is due to the variance in the velocity (which is minimised at $w_m$, where $\partial V/\partial w=0$), since the populations with higher velocities have longer radial decay lengths. As for the statistically distributed amplitude case, the shape of the filament can affect the results. In particular, a symmetric $\Lambda(x,w)$ leads to qualitatively similar results at low $w_*$, while at higher values the flattening is still present but very weak. 

\section{Discussion and Conclusions}

In this work, we discussed a number of different models for the filament evolution in the SOL and we applied them to a recently developed statistical framework which relates the fluctuations to the mean SOL profiles. The main strength of the statistical framework discussed in Sec.\ref{sec1.0} is its flexibility and its ability to be used with different models for the filament dynamics. It provides a rigorous basis to test their consistency with respect to experimental results but it also allows interpretation of the profiles through filamentary dynamics only. The purpose of this analysis is to shed some light on the non-exponential nature of the SOL profiles and on their response to different plasma conditions. While the models used here provide a simplified version of the actual filament dynamics, they should be able to capture physically motivated effects, such as the change in the exhaust timescale as the filaments evolve or their acceleration through weaker backgrounds, and to relate them to flattening and broadening of the SOL profiles. On the other hand, as better filamentary models will be developed or improved experimental measurements will become available, the conclusions of this paper could still be updated within the statistical framework.  

A number of mechanisms described in this paper might explain the flattening of the density and electron temperature in the far SOL. One of them is filament acceleration, which was observed in \cite{Zweben2016}. However, flattened profiles are observed also in machines like MAST \cite{Militello2016}, where such an acceleration is absent or very limited \cite{Kirk2016}. An increasing density exhaust timescale has theoretical justifications, as discussed in Sec.\ref{sec2.2}. For the density, this would require parallel gradients in the filaments which, while never directly observed, is quite reasonable \cite{Ham2016}. Indeed, if the filaments are ejected from the core and are not local instabilities in the SOL, a natural ballooning would occur due to their initial extension from X-point to X-point \cite{Kirk2006}. If they are born as linear instabilities in the SOL, theoretical results \cite{Myra1997,Militello2014} suggest that the varying curvature along the field lines would provide parallel gradients. As far as the electron temperature is concerned, a changing timescale arises due to the different mechanisms regulating the gradient removal and the sheath exhaust, but it could also occur because the cooling of the plasma triggers less efficient exhaust through the temperature dependent parallel heat conductivity. On the other hand, changing timescales lead to a radial decrease of the relative fluctuation amplitude, which is in contrast with experimental observations. Some other resilient mechanism should compensate for this. In this respect, the statistical distribution of initial amplitudes and perpendicular widths of the filaments, provides a flattening and an increase of the fluctuation amplitude and therefore seem to be an interesting candidate that could also explain the robustness of the flattening observations. In general, the non-exponential nature of the mean profiles and an increasing relative fluctuation amplitude seem to be a quite resilient feature in most of our models, thus putting these features of the SOL on a solid theoretical basis. 

As far as the density broadening is concerned, larger filament velocities do increase the decay length of the SOL, but also reduce the separatrix value of the density (the SOL confiememt worsens), which is not experimentally observed. In addition, this mechanism would also affect the temperature profiles which, however, do not seem to undergo a similar transition at high fuelling levels \cite{Lipschultz2005}. An interesting observation is that the broadening could occur if the SOL was `clogged' by neural particles that would increase the timescale for the particle exhaust through charge exchange interactions (see Section \ref{sec2.3}). This effect might be localised at the divertor, where the neutrals are denser due to the colder plasma temperature. This mechanism would not affect the electron temperature but it would cool down the ion temperature, which could be directly testable in experiments (ion temperature is, however, difficult to measure in the SOL). For the same reason, increased ionisation, which cools down the electrons, does not seem to be a strong candidate to explain the broadening although it might contribute to reduce the exhaust timescale (but a negative $\tau$, corresponding to more ionisation than exhaust, would lead to growing rather than decaying SOL profiles). It is also possible that filaments might experience a strong acceleration when at high fuelling level, possibly because they experience a larger resistivity as they move towards colder regions of the plasma \cite{Easy2016}. Finally, changes in the filament statistics lead to increases in the decay length which are modest in the case of width dependent velocity, and probably not strong enough to explain the broadening. A change in the mean initial amplitude has a more significant impact and might be compatible with the higher fuelling levels associated with the phenomenon. However, this would require significant $\eta_*$ and therefore velocity increases (of one order of magnitude) to produce visible effects.    

From our analysis, the flattening and the density of the SOL profiles emerge as complicated phenomena, which are likely to depend on several mechanisms at the same time. The statistics of the filament population plays an important role in determining profiles, so that approaches based on mean filament properties can miss dominant effects. Extrapolation of the observations in present-day machines is therefore delicate and would require first principle or at least empirical understanding of all the actors. This, in our opinion, motivates more theoretical and experimental investigations of filamentary dynamics and how it is modified by changes in the main plasma conditions. 

\acknowledgements

F.M. acknowledges useful discussions with Dr. Andrew Kirk, Dr. N. Walkden, Dr. F. Parra, Mr. T. Farley and Mr. L. Easy. We also thank Dr. Chris Ham for carefully reading the manuscript. This work has received funding from the RCUK Energy Programme [grant number EP/I501045]. To obtain further information on the data and models underlying this paper please contact PublicationsManager@ccfe.ac.uk. 

\appendix

\section{Analytic calculations}

From Eq.\ref{4b} and the assumptions of Sec.\ref{sec3.1} we write:
\begin{equation}
\label{13}
\Theta(x) = \frac{\eta_*}{\tau_w} \int_{-\infty}^{\infty}dtF(t)\Lambda\left(x-\int_0^t V(t')dt',w_*\right),
\end{equation}
which leads to:
\begin{equation}
\label{14}
\Theta(x) = \frac{\eta_*e^{\frac{x}{w_*}}}{\tau_w} \int_{-\infty}^{\infty}dt e^{-\frac{t}{\tau}-\frac{X(t)}{w_*}}H[X(t)-x]=\frac{\eta_*e^{\frac{x}{w_*}}}{\tau_w} \int_{X^{-1}(x)}^{\infty}dt e^{-\frac{t}{\tau}-\frac{X(t)}{w_*}}.
\end{equation}
From here, with constant filament velocity, $X=Vt$, it is easy to derive Eq.\ref{14b} and Eq.\ref{14c}.

\end{document}